\documentclass[11pt,draftcls,onecolumn,cha]{revtex4}

\usepackage{graphicx, subfigure}
\usepackage{amssymb}
\usepackage{amsthm}
\usepackage{graphicx}
\usepackage{dcolumn}
\usepackage{bm}

\begin{document}

\preprint{APS/000-QED}

\title{A chaotic system with only one stable equilibrium\thanks{}}

\author{Xiong Wang\footnote{Corresponding author.}}
\email{wangxiong8686@gmail.com}
 %\altaffiliation{Department of Electronic Engineering,
 %City University of Hong Kong, Hong Kong SAR, China.}
\author{Guanrong Chen}
\email{eegchen@cityu.edu.hk}
\affiliation{Department of Electronic
Engineering, City University of Hong Kong, Hong Kong SAR, China.}

\begin{abstract}
If you are given a simple three-dimensional autonomous quadratic
system that has only one stable equilibrium, what would you
predict its dynamics to be, stable or periodic? Will it be
surprising if you are shown that such a system is actually
chaotic? Although chaos theory for three-dimensional autonomous
systems has been intensively and extensively studied since the
time of Lorenz in the 1960s, and the theory has become quite
mature today, it seems that no one would anticipate a possibility
of finding a three-dimensional autonomous quadratic chaotic system
with only one stable equilibrium. The discovery of the new system,
to be reported in this Letter, is indeed striking because for a
three-dimensional autonomous quadratic system with a single stable
node-focus equilibrium, one typically would anticipate non-chaotic
and even asymptotically converging behaviors. Although the new
system is of non-hyperbolic type, therefore the familiar
\v{S}i'lnikov homoclinic criterion is not applicable, it is
demonstrated to be chaotic in the sense of having a positive
largest Lyapunov exponent, a fractional dimension, a continuous
broad frequency spectrum, and a period-doubling route to chaos.
\end{abstract}

%\pacs{}

\maketitle

\section{Introduction}

For three-dimensional (3D) autonomous hyperbolic type of chaotic
systems, a commonly accepted criterion for proving the existence
of chaos is due to \v{S}i'lnikov [1-4], which has a slight
extension recently [5]. Chaos in the \v{S}i'lnikov type of 3D
autonomous quadratic dynamical systems may be classified into four
subclasses [6]:

$\bullet$ chaos of the \v{S}i'lnikov homoclinic-orbit type;

$\bullet$ chaos of the \v{S}i'lnikov heteroclinic-orbit type;

$\bullet$ chaos of the hybrid type with both \v{S}i'lnikov
homoclinic and heteroclinic orbits;

$\bullet$ chaos of other types.

In this classification, a system is required to have a
saddle-focus type of equilibrium, which belongs to the hyperbolic
type at large.

Notice that although most chaotic systems are of hyperbolic type,
there are still many others that are not so. For non-hyperbolic
type of chaos, saddle-focus equilibrium typically does not exist
in the systems, as can be seen from Table I which includes several
non-hyperbolic chaotic systems found by Sprott [7-10]. More
recently, Yang and Chen also found a chaotic system with one
saddle and two stable node-foci [11] and, moreover, an unusual 3D
autonomous quadratic Lorenz-like chaotic system with only two
stable node-foci [12]. In fact, similar examples can be easily
found from the literature.

\begin{table}[h]
\caption{Equilibria and eigenvalues of several typical Sprott
systems.}
{\begin{tabular}{c c c c}\\[-2pt]
\hline
Systems & Equations & Equilibria & Eigenvalues \\[6pt]
\hline\\[-5pt]

Sprott& $\dot x=-y$&$(0,0,0)$ &$0, \pm i$ \\[1pt]
Case D &$\dot y=x+z$&{}&{}\\[2pt]
{} &$\dot z=xz+3y^2$ &{} &{} \\\hline\\[-5pt]

Sprott& $\dot x=yz$&$(0.25, 0.0625, 0)$ &$-1, \pm 0.5i$ \\[1pt]
Case E &$\dot y=x^2-y$&{}&{}\\[2pt]
{} &$\dot z=1-4x$ &{} &{} \\\hline\\[-5pt]

Sprott& $\dot x=-0.2y$&$(0, 0, 0)$ &$-1.13449, 0.06725\pm 0.58996i$ \\[1pt]
Case I &$\dot y=x+z$&{}&{}\\[2pt]
{} &$\dot z=x+y^2-z$ &{} &{} \\\hline\\[-5pt]

Sprott& $\dot x=2z$&$(0, 0, 0)$ &$-2.31460, 0.15730\pm 1.30515i$ \\[1pt]
Case J &$\dot y=-2y+z$&{}&{}\\[2pt]
{} &$\dot z=-x+y+y^2$ &{} &{} \\\hline\\[-5pt]

Sprott& $\dot x=y+3.9z$&$(1, 0.9, -0.23077)$ &$-1.43329,
0.21664\pm 1.63526i$ \\[1pt]
Case L &$\dot y=0.9x^2-y$&{}&{}\\[2pt]
{} &$\dot z=1-x$ &{} &{} \\\hline\\[-5pt]

Sprott& $\dot x=-2y$&$(-0.25, 0, 0.5)$ &$-2.31460, 0.15730\pm 1.30515i$ \\[1pt]
Case N &$\dot y=x+z^2$&{}&{}\\[2pt]
{} &$\dot z=1+y-2z$ &{} &{} \\\hline\\[-5pt]

Sprott& $\dot x=0.9-y$&$(-0.44444, 0.9, -0.4)$ &$-1.23212,
0.11606\pm 0.84674i$ \\[1pt]
Case R &$\dot y=0.4+z$&{}&{}\\[2pt]
{} &$\dot z=xy-z$ &{} &{} \\

\hline
\end{tabular}}
\end{table}

In this paper, we report a very surprising finding of a simple 3D
autonomous chaotic system that has only one equilibrium and,
furthermore, this equilibrium is a stable node-focus. For such a
system, one almost surely would expect asymptotically convergent
behaviors or, at best, would not anticipate chaos per se.

From Table I, one may observe that the Sprott D and E systems also
have only one equilibrium, but nevertheless this equilibrium is
not stable. From this point of view, it is easy to understand and
indeed easy to prove that the new system will not be topologically
equivalent to the Sprott systems.

\section{The new system}

\subsection{The mechanism of generating the new system}

The mechanism of generating the new system is simple and
intuitive.

To start with, let us first review some of the Sprott chaotic
systems listed in Table I, namely those with only one equilibrium.
One can easily see that systems I, J, L, N and R all have only one
saddle-focus equilibrium, while systems D and E both degenerate in
the sense that their Jacobian eigenvalues at the equilibria
consist of one conjugate pair of pure imaginary numbers and one
real number. Clearly, the corresponding equilibria are not stable.

It is also easy to imagine that a tiny perturbation to the system
may be able to change such a degenerate equilibrium to a stable
one. Therefore, we added a simple constant control parameter to an
aforementioned Sprott chaotic system, trying to change the
stability of its single equilibrium to a stable one while
preserving its chaotic dynamics.

As a result, we obtained the following new system:
\begin{equation}\label{wangeq}
\left\{
\begin{array}{l}
\dot{x}=yz+a\\
\dot{y}=x^{2}-y \\
\dot{z}=1-4x.
\end{array}
\right.
\end{equation}

When $a=0$, it is the Sprott E system; when $a\neq0$, however, the
stability of the single equilibrium is fundamentally different, as
can be verified and compared between the results shown in Table I
and Table II, respectively.

\begin{table}[h]
\caption{Equilibria and eigenvalues of the new system.}
{\begin{tabular}{c c c c}\\[-2pt]
\hline
Systems & Equations & Equilibria & Eigenvalues \\[6pt]
\hline\\[-5pt]

New System& $\dot x=yz+a$&$(0.25, 0.0625, 0.08)$ &$-1.03140, 0.01570\pm 0.49208i$ \\[1pt]
a=-0.005 &$\dot y=x^2-y$&{}&{}\\[2pt]
{} &$\dot z=1-4x$ &{} &{} \\\hline\\[-5pt]

New System& $\dot x=yz+a$&$(0.25, 0.0625, -0.096)$ &$-0.96069, -0.01966\pm 0.50975i$ \\[1pt]
a=0.006 &$\dot y=x^2-y$&{}&{}\\[2pt]
{} &$\dot z=1-4x$ &{} &{} \\\hline\\[-5pt]

New System& $\dot x=yz+a$&$(0.25, 0.0625, -0.352)$ &$-0.84580, -0.07710\pm 0.53818i$ \\[1pt]
a=0.022 &$\dot y=x^2-y$&{}&{}\\[2pt]
{} &$\dot z=1-4x$ &{} &{} \\\hline\\[-5pt]

New System& $\dot x=yz+a$&$(0.25, 0.0625, -0.48)$ &$-0.78217, -0.10891\pm 0.55476i$ \\[1pt]
a=0.030 &$\dot y=x^2-y$&{}&{}\\[2pt]
{} &$\dot z=1-4x$ &{} &{} \\\hline\\[-5pt]

New System& $\dot x=yz+a$&$(0.25, 0.0625, -0.8)$ &$-0.60746, -0.19627\pm 0.61076i$ \\[1pt]
a=0.050 &$\dot y=x^2-y$&{}&{}\\[2pt]
{} &$\dot z=1-4x$ &{} &{} \\[-2pt] \hline

\end{tabular}}
\end{table}

To better understand the new system (\ref{wangeq}), and more
importantly to demonstrate that this new system is indeed chaotic,
some basic properties of the system are briefly analyzed next.

\subsection{Equilibrium and stability}

The system (\ref{wangeq}) possesses only one equilibrium:
\begin{equation}
P \left( x_{E},y_{E},z_{E} \right)
=\left(\frac{1}{4},\frac{1}{16},-16\,a\right).
\end{equation}

Linearizing the system at the equilibrium $P$ gives the Jacobian
matrix
\begin{equation}
J=\left[
\begin {array}{ccc} 0&z&y\\ \noalign{\medskip}2\,x&-1&0
\\ \noalign{\medskip}-4&0&0\end {array} \right] = \left[
\begin {array}{ccc} 0&-16\,a&\frac{1}{16}\\
\noalign{\medskip}\frac{1}{2}&-1&0
\\ \noalign{\medskip}-4&0&0\end {array}
 \right].
\end{equation}
By solving the characteristic equation $|\lambda I - J|=0$, one
obtains the Jacobian eigenvalues, as shown in Table II for some
chosen values of the parameter $a$.

\subsection{Lyapunov exponents}

To verify the chaoticity of system (\ref{wangeq}), its Lyapunov
exponents and Lyapunov dimension are calculated.

The Lyapunov exponents are denoted by $L_i$, $i=1,2,3$, and
ordered as $L_1>L_2>L_3$. A system is considered chaotic if
$L_1>0,\,L_2=0,\,L_3<0$ with $|L_1|<|L_3|$.

The Lyapunov dimension is defined by
\[
D_L=j+\frac{1}{|L_{j+1}|}\sum_{i=1}^j{L_i},
\]
where $j$ is the largest integer satisfying
$\sum_{i=1}^j{L_i}\ge0$ and $\sum_{i=1}^{j+1}{L_i}<0$.

\begin{figure*}
\centering \subfigure{
\begin{minipage}[b]{0.7\textwidth}
\includegraphics[width=12cm]{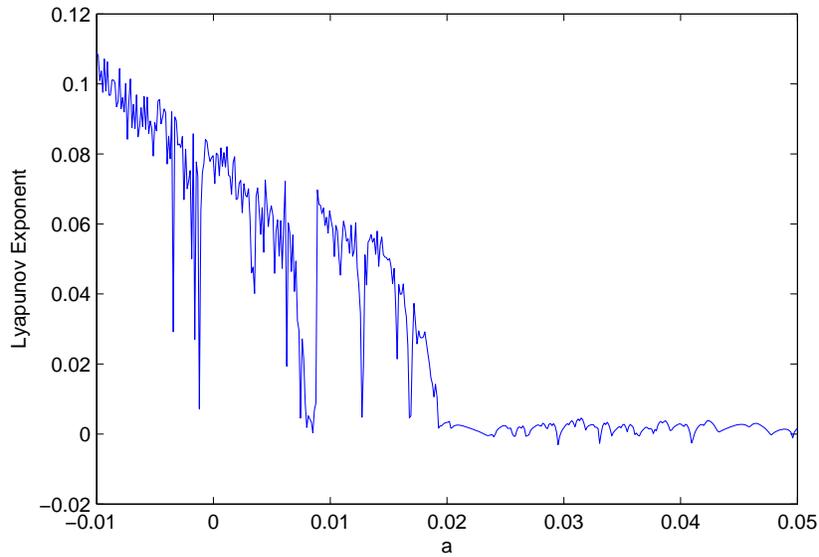}
\end{minipage}}
\caption{(Color online) The largest Lyapunov exponent versus the
parameter $a$.} \label{Lya}
\end{figure*}

FIG. \ref{Lya} shows the dependence of the largest Lyapunov
exponent of system (\ref{wangeq}) on the parameter $a$. From
FIG. \ref{Lya}, it is clear that the largest Lyapunov exponent
decreases as the parameter $a$ increases from $-0.01$ to $0.05$.

\subsection{The degenerate case of $a=0$ (Sprott E system)}

When $a=-0.005$, the system equilibrium is of the regular
saddle-focus type; this case of the chaotic system has been
studied before therefore will not be discussed here.

\begin{figure*}
\centering \subfigure{
\begin{minipage}[b]{1\textwidth}
\includegraphics[width=16cm]{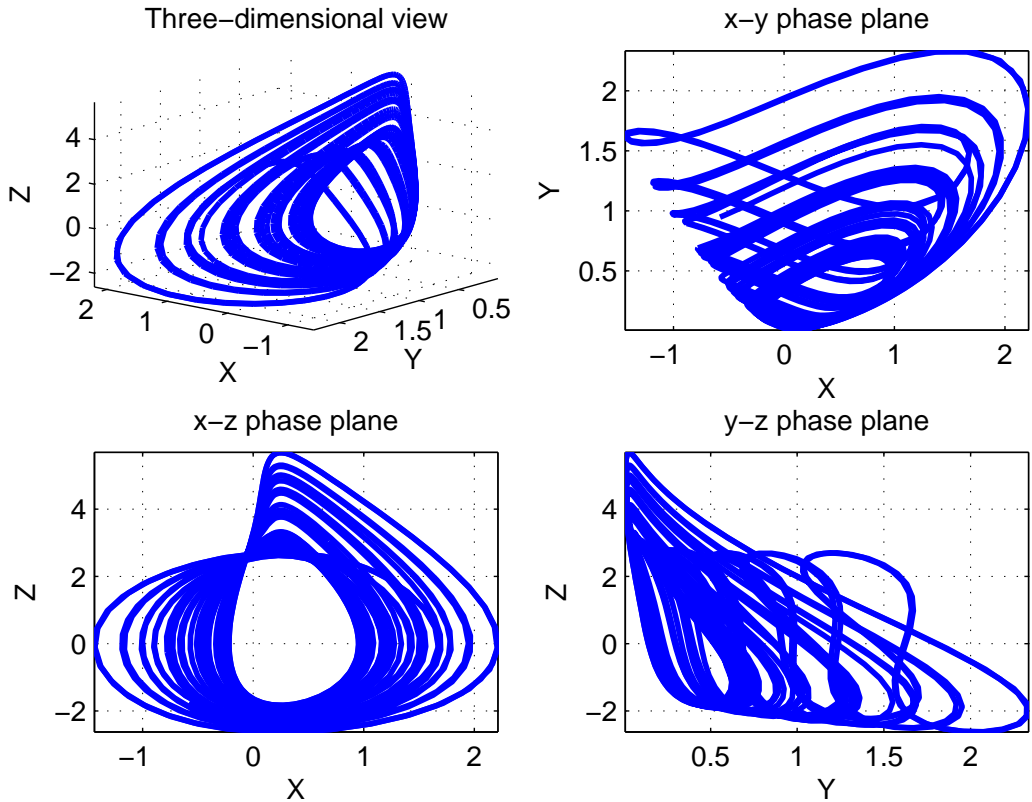}
\end{minipage}}
\caption{(Color online) The new system: chaotic attractor with
$a=0$, including 3D views on the $x$-$y$ plane, $x$-$z$ plane and
$y$-$z$ plane.} \label{0003D}
\end{figure*}

When $a=0$, the equilibrium degenerates. It is precisely the
Sprott E system listed in Table I (see FIG. \ref{0003D}). The
\v{S}i'lnikov homoclinic criterion might be applied to this system
to show the existence of chaos, however, but it involves somewhat
subtle mathematical arguments.

\begin{figure*}
\centering \subfigure{
\begin{minipage}[b]{1\textwidth}
\includegraphics[width=16cm]{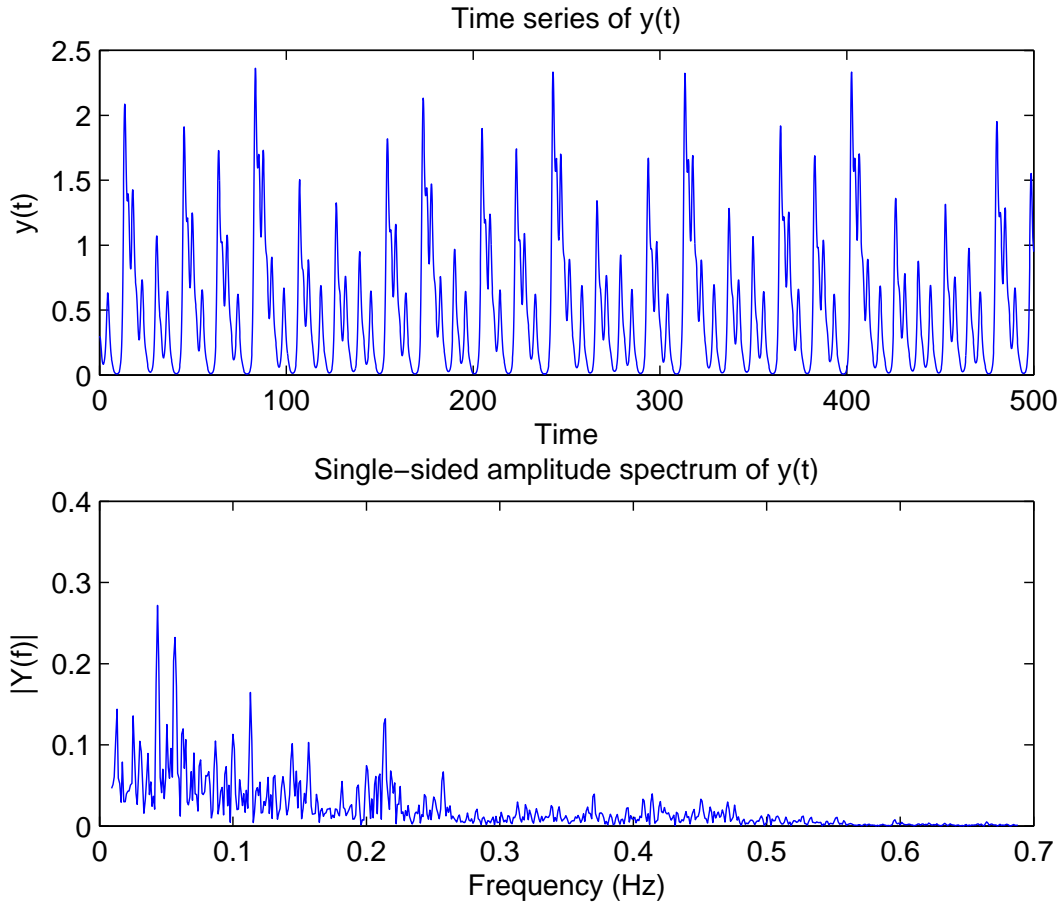}
\end{minipage}}
\caption{(Color online) Top: An apparently chaotic waveform of
$y(t)$ ($a=0$). Bottom: An apparently continuous broadband
frequency spectrum $|y(t)|$.}\label{000}
\end{figure*}

In this degenerate case, the positive largest Lyapunov exponent of
the system (see Table II) still indicates the existence of chaos.
In the time domain, FIG. \ref{000} (top part) shows an apparently
chaotic waveform of $y(t)$; while in the frequency domain, FIG.
\ref{000} (bottom part) shows an apparently continuous broadband
spectrum $|y(t)|$. These all prove that the Sprott E system, or
the new system (\ref{wangeq}) with $a=0$, is indeed chaotic.

\bigbreak

\subsection{The case of $a=0.006$: a new type of chaos}

When $a>0$, the stability of the equilibrium is fundamentally
different from that of the Sprott E system. In this case, the
equilibrium becomes  a node-focus (see Table II). The
\v{S}i'lnikov homoclinic criterion is therefore inapplicable to
this case.

Take $a=0.006$ as an example. Numerical calculation of the
Lyapunov exponents gives $L_1 = 0.0489$, $L_2 =0$ and $L_3
=-1.0485$, indicating the existence of chaos.

\begin{figure*}
\centering \subfigure{
\begin{minipage}[b]{1\textwidth}
\includegraphics[width=16cm]{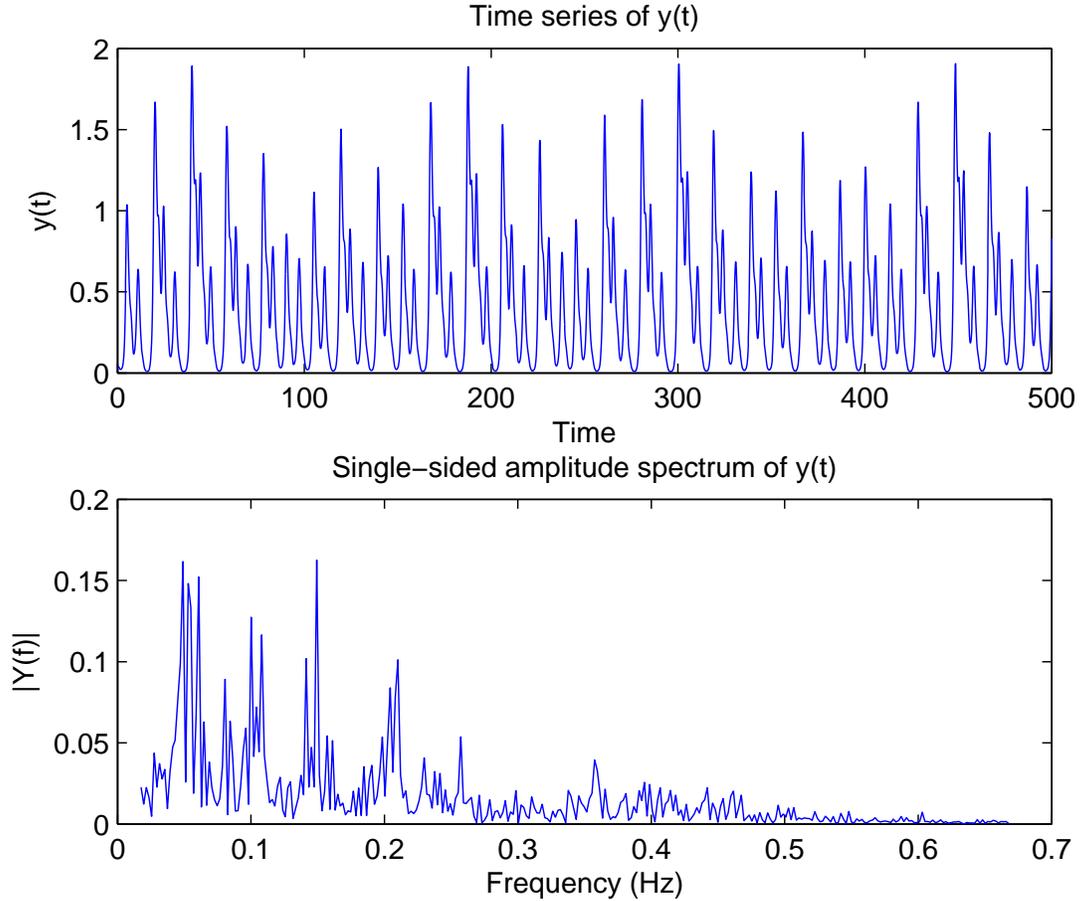}
\end{minipage}}
\caption{(Color online) Top: An apparently chaotic waveform of
$y(t)$ ($a=0.006$). Bottom: An apparently continuous broadband
frequency spectrum $|y(t)|$.} \label{006}
\end{figure*}

In the time domain, FIG. \ref{006} (top part) shows an apparently
chaotic waveform $y(t)$; while in the frequency domain,
FIG. \ref{006} (bottom part) shows an apparently continuous
broadband spectrum $|y(t)|$. These all prove that the new system
(\ref{wangeq}) with $a=0.006$ is indeed chaotic.

\begin{figure*}
\centering \subfigure{
\begin{minipage}[b]{1\textwidth}
\includegraphics[width=16cm]{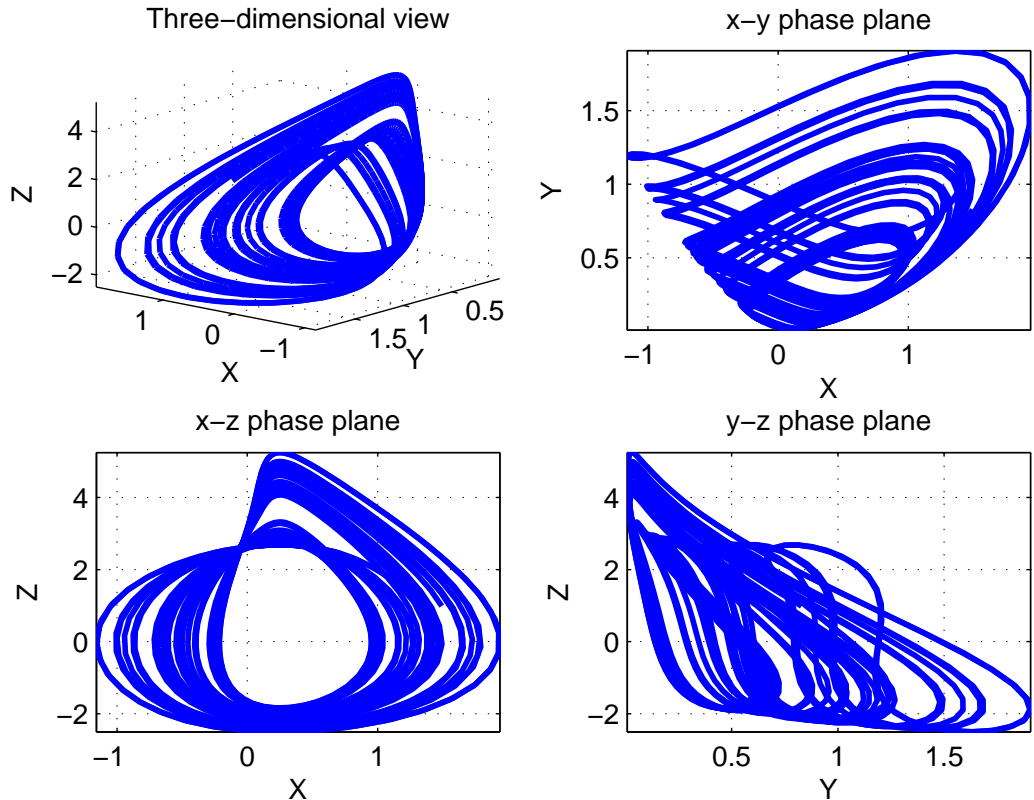}
\end{minipage}}
\caption{(Color online) The new system: chaotic attractor with
$a=0.006$, including 3D views on the $x$-$y$ plane, $x$-$z$ plane
and $y$-$z$ plane.} \label{0063D}
\end{figure*}

\bigbreak

\subsection{Bifurcations analysis}

\begin{figure*}
\centering
\includegraphics[width=8cm]{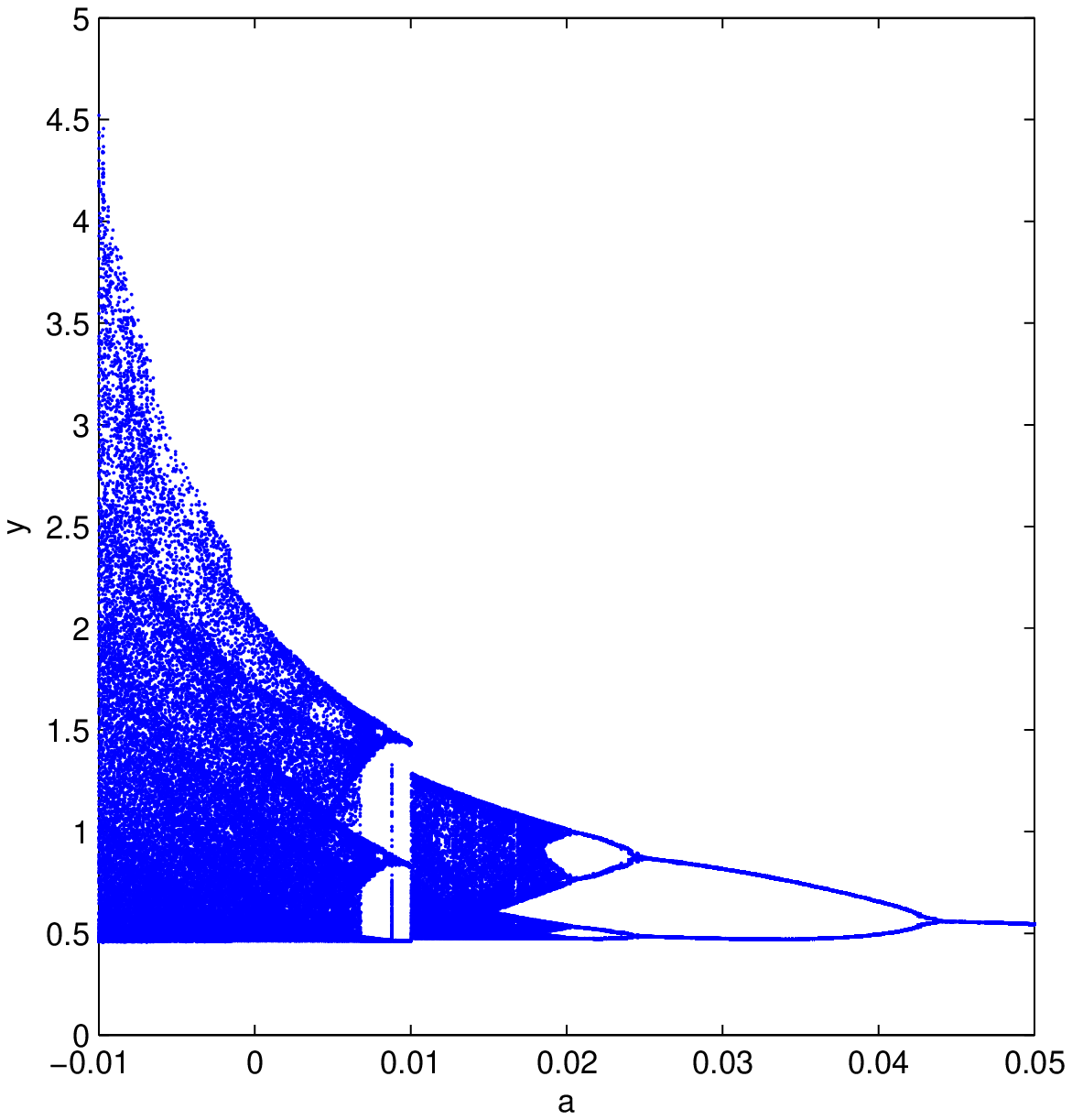}
\caption{(Color online) Bifurcation diagram, showing a
period-doubling route to chaos in $y$ (at $x=0.25$) versus the
parameter $a$.} \label{bif}
\end{figure*}

FIG. \ref{bif} shows a bifurcation diagram versus the parameter
$a$, demonstrating a period-doubling route to chaos.

\begin{figure}[h]
\begin{minipage}[t]{0.45\textwidth}
\centering
\includegraphics[width=\textwidth]{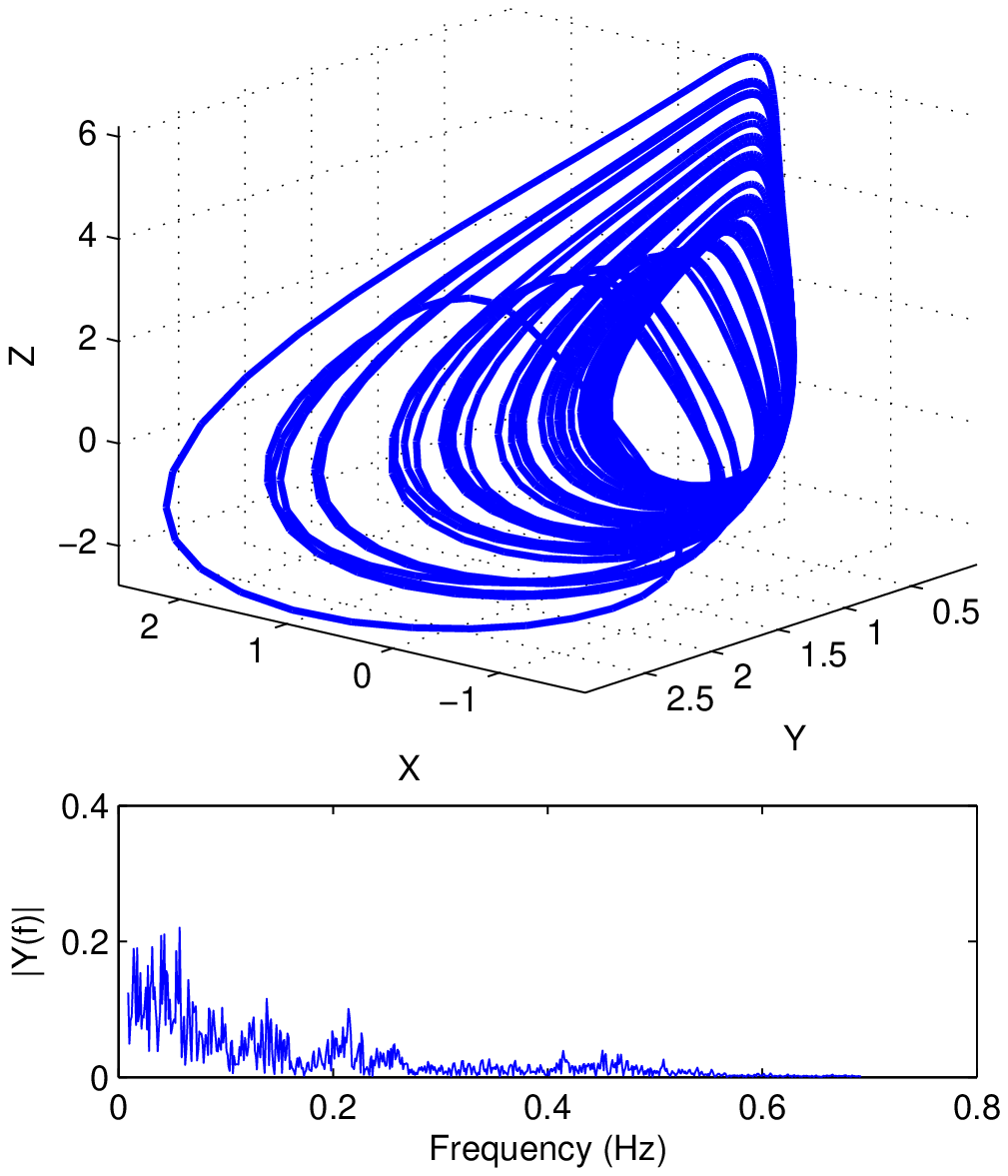}
{\small (a)}
\end{minipage}
\hfill
\begin{minipage}[t]{0.45\linewidth}
\centering
\includegraphics[width=\textwidth]{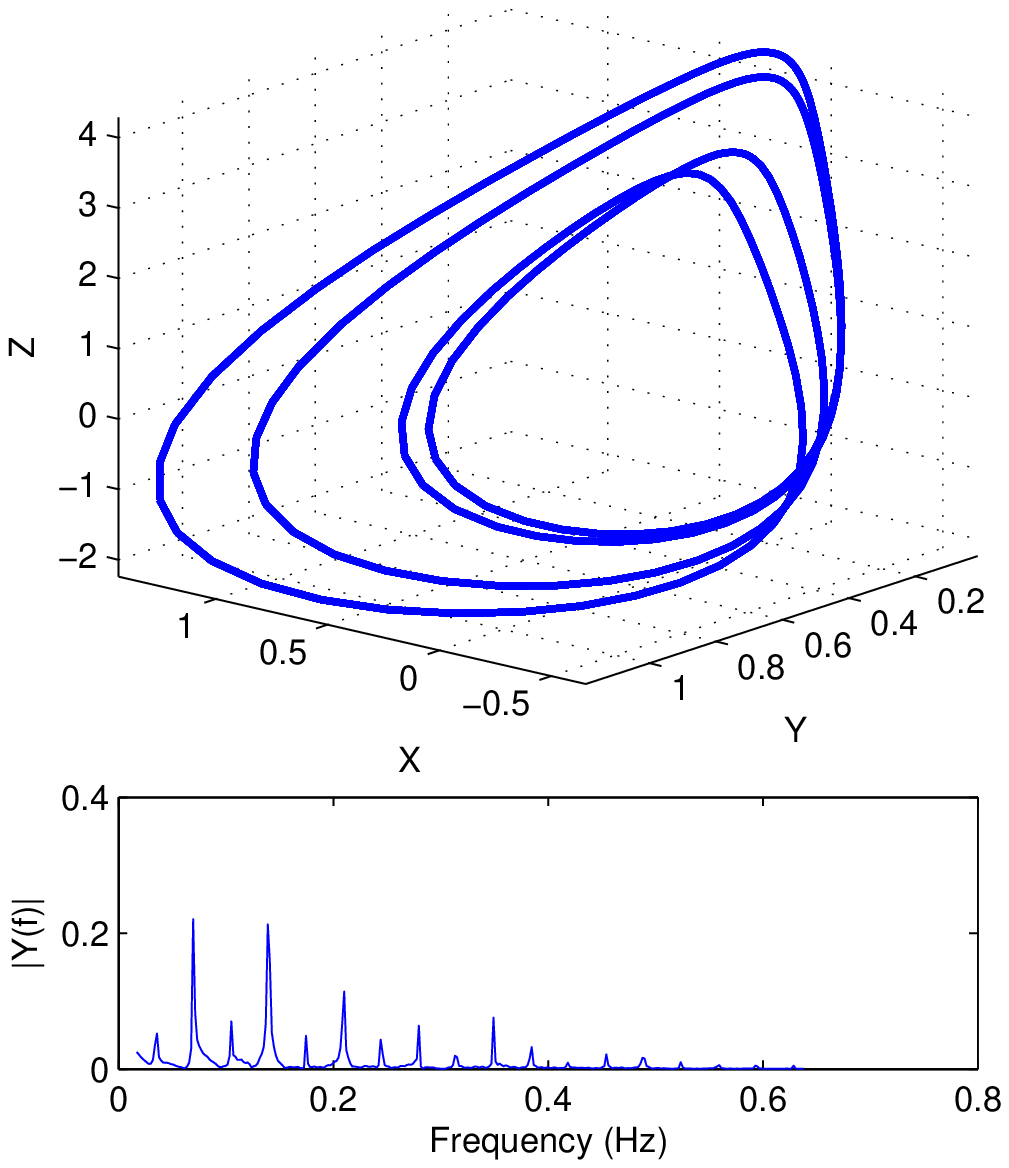}
{\small (b)}
\end{minipage}

\begin{minipage}[t]{0.45\linewidth}
\centering
\includegraphics[width=\textwidth]{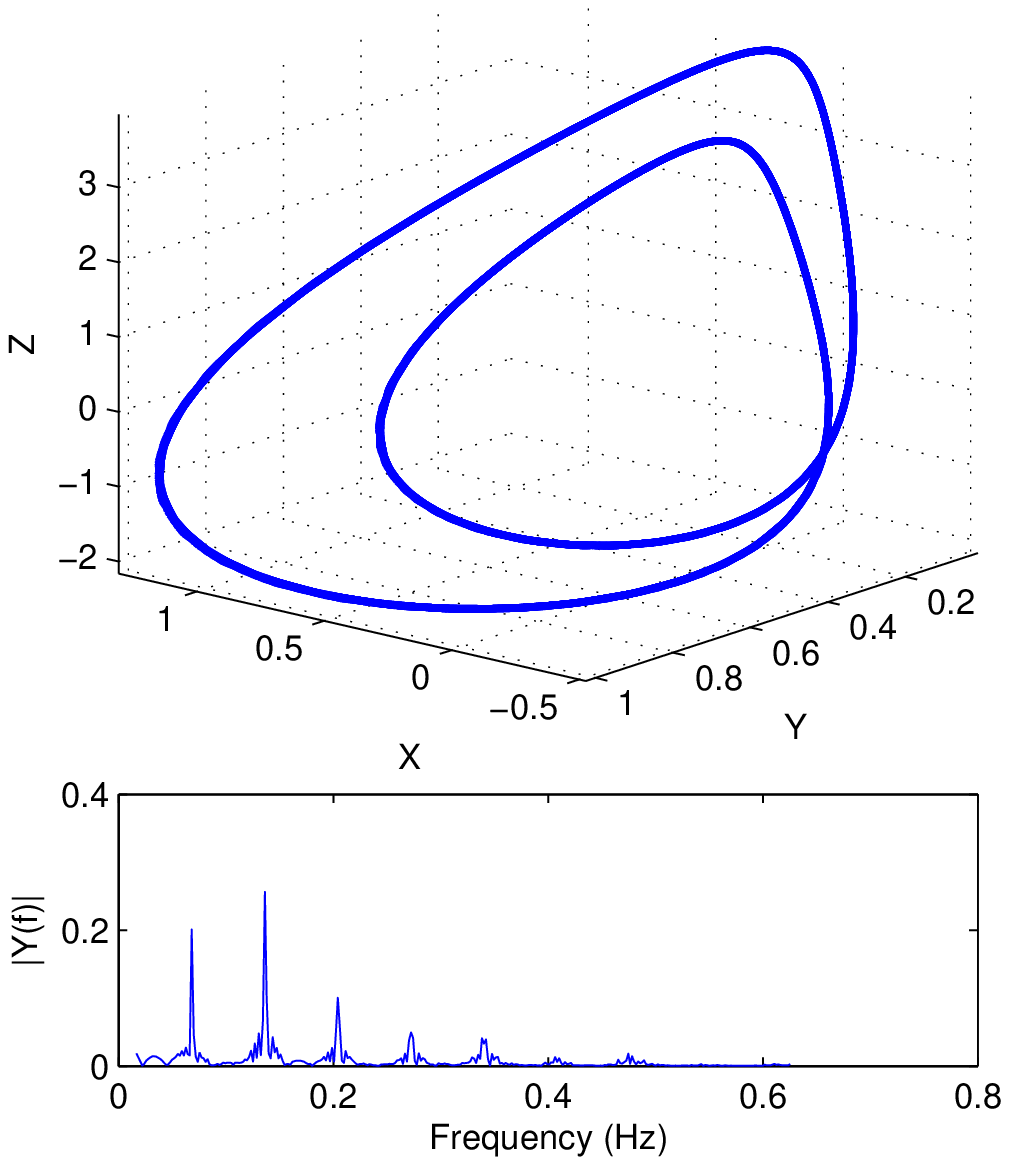}
{\small (c)}
\end{minipage}
\hfill
\begin{minipage}[t]{0.45\linewidth}
\centering
\includegraphics[width=\textwidth]{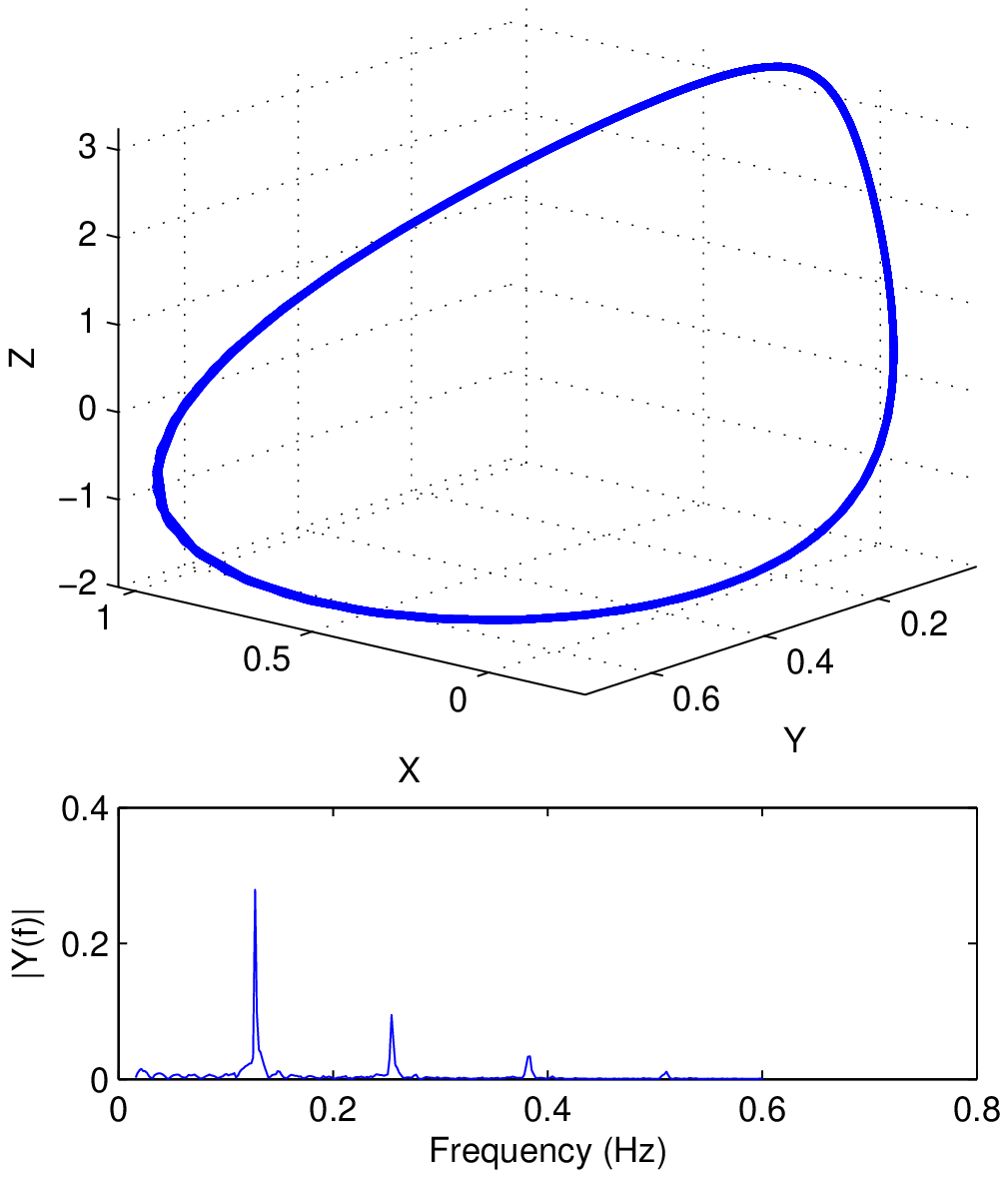}
{\small (d)}
\end{minipage}
\caption{(Color online) Phase portraits and frequency spectrums:
(a) $a=0.006$, (b) $a=0.022$, (c) $a=0.03$, (d) $a=0.05$.}
\label{bif2}
\end{figure}

FIG. \ref{bif2} also demonstrates the gradual evolving dynamical
process as $a$ is continuously varied.

Both figures indicate that although the equilibrium is changed
from an unstable saddle-focus to a stable node-focus, the chaotic
dynamics survive in a relative narrow range of the parameter $a$.

All the above numerical results are summarized in Table III.

\begin{table}[h]
\caption{Numerical results for some values of the parameter $a$
with initial values $(1,1,1)$.}
{\begin{tabular}{c c c c}\\[-2pt]
\hline

Parameters & Eigenvalues & Lyapunov Exponents & Fractal Dimensions \\[6pt]
\hline\\[-5pt]

$a=-0.005$ & $\lambda_1=-1.03140$& $L_1=0.0884$ &{}\\[1pt]
{} & $\lambda_{2,3}=0.01570\pm 0.49208i$ & $L_2=0$ & $D_L=2.081$ \\[2pt]
{} & {} & $L_3=-1.0884$ & {} \\\hline\\[-5pt]

$a=0$ & $\lambda_1=-1$ & $L_1=0.0766$ & {} \\[1pt]
{} & $\lambda_{2,3}=\pm 0.5i$ & $L_2=0$ & $D_L=2.071$ \\[2pt]
{} & {} & $L_3=-1.0766$ & {} \\\hline\\[-5pt]

$a=0.006$ & $\lambda_1=-0.96069$& $L_1=0.0510$ &{}\\[1pt]
{} & $\lambda_{2,3}=-0.01966\pm 0.50975i$ & $L_2=0$ & $D_L=2.048$ \\[2pt]
{} & {} & $L_3=-1.0510$ & {} \\\hline\\[-5pt]

$a=0.022$ & $\lambda_1=-0.84580$ & $L_1=0$ & {} \\[1pt]
{} & $\lambda_{2,3}=-0.07710\pm 0.53818i$ & $L_2=-0.1381$ & $D_L=1.000$ \\[2pt]
{} & {} & $L_3=-0.8619$ & {} \\\hline\\[-5pt]

$a=0.030$ & $\lambda_1=-0.78217$ & $L_1=0$ & {} \\[1pt]
{} & $\lambda_{2,3}=-0.10891\pm 0.55476i$ & $L_2=-0.0826$ & $D_L=1.000$ \\[2pt]
{} & {} & $L_3=-0.9174$ & {} \\\hline\\[-5pt]

$a=0.050$ & $\lambda_1=-0.60746$ & $L_1=0$ & {} \\[1pt]
{} & $\lambda_{2,3}=-0.19627\pm 0.61076i$ & $L_2=-0.0518$ & $D_L=1.001$ \\[2pt]
{} & {} & $L_3=-0.9482$ & {} \\[-5pt] \hline

\end{tabular}}
\end{table}

\section{Conclusion}

This paper has reported the finding of a simple three-dimensional
autonomous chaotic system which, very surprisingly, has only one
stable node-focus equilibrium. The discovery of this new system is
striking, because with a single stable equilibrium in a 3D
autonomous quadratic system, one typically would anticipate
non-chaotic and even asymptotically converging behaviors. Yet,
unexpectedly, this system is chaotic. Although the new system is
non-hyperbolic type, therefore the \v{S}i'lnikov homoclinic
criterion is not applicable, it has been verified to be chaotic in
the sense of having a positive largest Lyapunov exponent, a
fractional dimension, a continuous frequency spectrum, and a
period-doubling route to chaos.

Although the fundamental chaos theory for autonomous dynamical
systems seems to have reached its maturity today, our finding
reveals some new mysterious features of chaos.

\subsection*{Acknowledgement}

This research was supported by the National Natural Science
Foundation of China under grant 10832006 and the Hong Kong
Research Grants Council under grant CityU1117/10E.

%\section*{References}

\end{document}